\documentclass[prl,twocolumn, aps,amssymb,footinbib,showpacs]{revtex4-1}

\usepackage{graphicx}
\usepackage{amsmath,amssymb}
\usepackage{cancel}
\usepackage{color}

\newcommand{\ua}{\uparrow}
\newcommand{\da}{\downarrow}

\begin{document}

\title{Enhanced anti-ferromagnetic exchange between magnetic impurities in a superconducting host}

\author{N. Y. Yao$^{1}$, L. I. Glazman$^{2,3}$,  E. A. Demler$^{1}$, M. D. Lukin$^{1}$, J. D. Sau$^{4}$ }
\affiliation{$^{1}$Department of Physics, Harvard University, Cambridge, MA 02138, U.S.A.}
\affiliation{$^{2}$Department of Physics, Yale University, New Haven, CT 06520, U.S.A.}
\affiliation{$^{3}$Department of Applied Physics, Yale University, New Haven, CT 06520, U.S.A.}
\affiliation{$^{4}$Joint Quantum Institute and Condensed Matter Theory Center, Department of Physics,
University of Maryland, College Park, Maryland 20742, U.S.A.}
\begin{abstract}

It is generally believed that superconductivity only weakly affects the indirect exchange between magnetic impurities. If the distance $r$ between impurities is smaller than  than the superconducting coherence length ($r\lesssim\xi$), this exchange is thought to be dominated by RKKY interactions, identical to the those in a normal metallic host. This perception is based on a perturbative treatment of the exchange interaction. Here, we provide a  non-perturbative analysis and demonstrate that the presence of Yu-Shiba-Rusinov bound states induces a strong $1/r^2$ anti-ferromagnetic interaction that can dominate over conventional RKKY even at distances significantly smaller than the  coherence length ($r\ll\xi$).  Experimental signatures, implications and applications are discussed.

\end{abstract}
\maketitle

Understanding the interactions between magnetic impurities (localized spins) in a metallic host represents an important question at the interface of fundamental and applied science \cite{Vleck62, Ruderman54, Kasuya56, Yosida57, Balatsky06}. 
While spins always interact with one another via their intrinsic dipolar interaction, in a metal, their mutual interaction with conduction electrons can significantly enhance the effective interactions. For simple metals, this  results in the so-called RKKY (Ruderman-Kittel-Kasuya-Yosida) interaction \cite{Ruderman54, Kasuya56, Yosida57} ---  a coupling mechanism between magnetic moments in which one impurity partially polarizes the spin of  conduction electrons; the second impurity then interacts with the spin density of the itinerant electrons, thereby inducing an effective long-range  interaction. One of the crucial predictions of RKKY is the oscillatory sign of the exchange interaction, a feature which underlies giant magnetoresistance \cite{Baibich88,Binasch89}.

More recently,  significant effort has been devoted to understanding magnetic impurities on the surface of superconducting metals \cite{Balatsky06,Moca08,Poilblanc94,Hatte97a,Hatte97b,Flatte00,Balatsky95,Salkola97,Pan00,Fominov11}. 
This owes in part, to experimental advances in single adatom control, which have enabled the observation of locally modified electronic properties and raise the tantalizing prospect of atom-by-atom construction of magnetic nanostructures \cite{Yazdani97, Ji08,Iavarone10}. 
Moreover, interactions between such impurities may play a role in explaining low-frequency flux noise in Josephson circuits \cite{Faoro08,Sendelbach08}.
%
%

\begin{figure}
\includegraphics[width=3.4in]{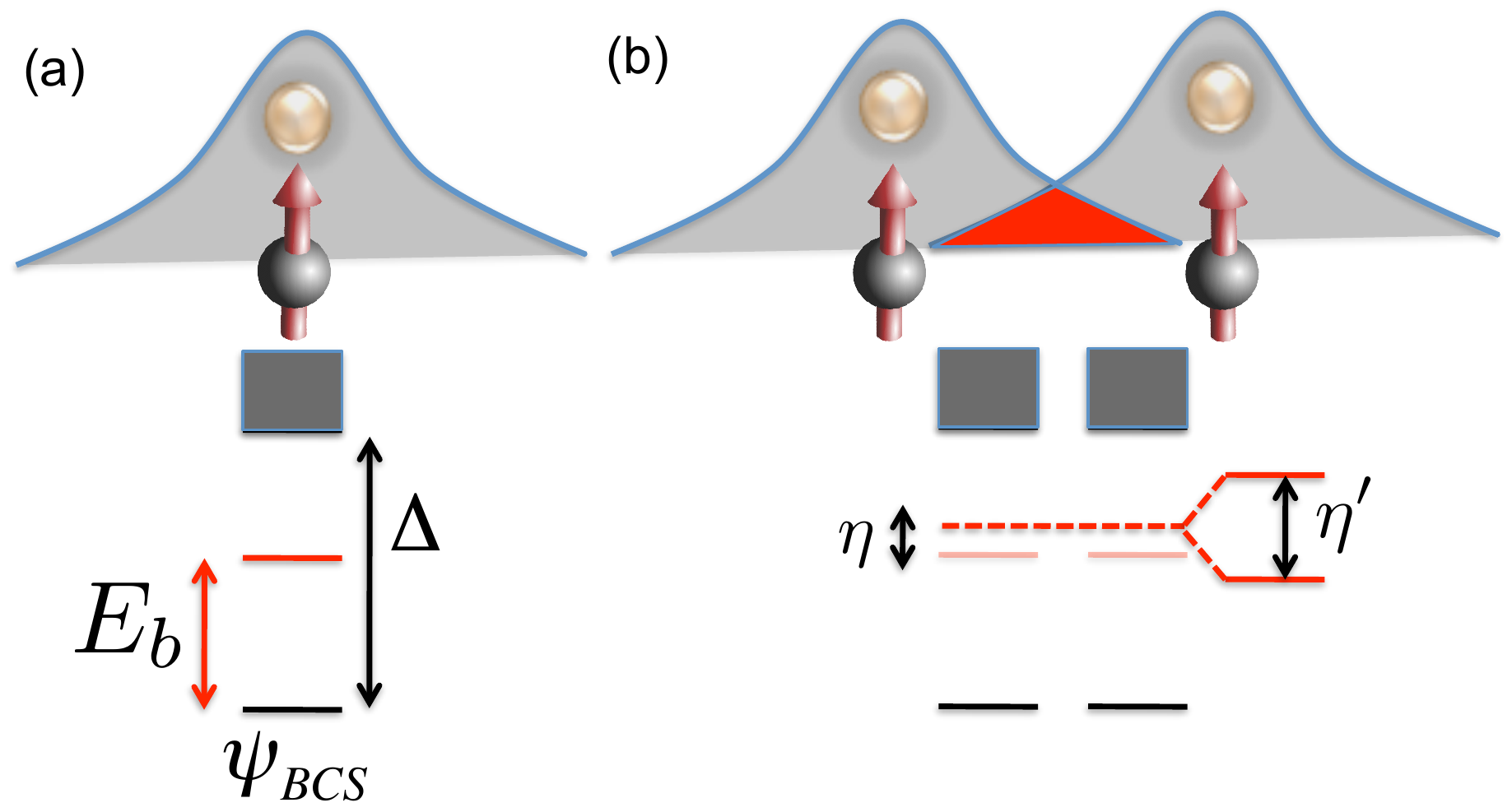}
\caption{a) Schematic illustration of a magnetic impurity which binds a localized electronic YSR state. The associated spectrum is shown below, with the BCS ground state $\psi_{BCS}$ separated from excited states by $\Delta$. There exists a single mid-gap YSR state of energy $E_{b}$. b) When two impurities are separated by distances $r < \xi$, their YSR states overlap and hybridize. This hybridization causes both an overall energy shift $\eta$ and a splitting $\eta'$.
 }
\label{fig:YSRscheme}
\end{figure}

The effect of superconductivity on RKKY interactions is well established at lowest-order perturbation theory (Born approximation) in the exchange interaction between the localized and itinerant spins. 
In particular, the suppressed spin susceptibility in the superconducting ground state modifies the inter-impurity interaction to become purely anti-ferromagnetic when the separation between the impurities exceeds the superconducting coherence length ($r\gtrsim\xi$); at such distances however, the strength of this antiferromagnetic exchange is exponentially small in the separation $r$. On the other hand, for impurities separated by distances $r < \xi$,  conventional RKKY dominates  the effective interaction and  superconductivity yields only a weak antiferromagnetic correction  \cite{Galitski02, Aristov97, Abrikosov88}. 
Crucially, this perturbative treatment neglects the formation of so-called  Yu-Shiba-Rusinov (YSR) bound states---localized electronic states that arise near a  magnetic impurity.

In this Letter, we show that by tuning the energy of YSR states close to the middle of the superconducting gap, one may substantially enhance the antiferromagnetic contribution stemming from  the indirect spin exchange, allowing it to dominate over conventional RKKY even at distances $r\lesssim \xi$ \cite{Yu65, Shiba68, Rusinov69}. When two magnetic impurities are brought near one another, their associated YSR states hybridize in a spin-dependent fashion, yielding an effective interaction. That one might expect such an interaction to dominate over  RKKY results, in part,  from the strong localization of the YSR state around the impurity, directly contrasting with the delocalized scattering states that mediate RKKY.  This localization implies that quasiparticles bound to the YSR states are more strongly coupled to the impurity and therefore might be expected to mediate stronger exchange.



The key ideas underlying our derivation are illustrated in Fig.~1. We begin by considering a BCS superconductor with Hamiltonian,
\begin{equation}
H_{0} = \sum_{\text{\bf k}, \sigma} \epsilon_\text{\bf k} c_{\text{\bf k},\sigma}^{\dagger} c_{\text{\bf k},\sigma} + \Delta \sum_\text{\bf k} [ c_{\text{\bf k} \uparrow}^{\dagger} c_{-\text{\bf k} \downarrow}^{\dagger} + c_{-\text{\bf k} \downarrow} c_{\text{\bf k} \uparrow}].
\end{equation}
The associated spectrum (Fig.~1) depicts the BCS ground state, $\psi_{BCS}$, separated from excited states by the superconducting gap $\Delta$. 
In the presence of a spin impurity whose contact exchange interaction is of strength $J$, an excited-state electron can lower its energy below the superconducting gap by aligning its spin opposite the direction of the impurity. Treating the spin impurity classically yields the existence of a localized  bound state (YSR state) of energy \cite{Yu65, Shiba68, Rusinov69},
\begin{equation}
E_{b} = \Delta \frac{1-(\pi J S N_0 /2)^2}{1+(\pi J S N_0 /2)^2} = \Delta \frac{1-\beta^2}{1+\beta^2}
\end{equation}
where $N_0$ is the normal state DOS at the Fermi energy. For pure exchange scattering, the YSR energy is conveniently re-expressed in terms of a phase shift $\tan(\delta)  \equiv \beta = \pi J S N_0 / 2$, wherein $E_{b} = \Delta \cos (2\delta)$.  The latter relation between $E_{b}$ and $\delta$ is more general than Eq.~(2), and valid beyond the approximation of a classical magnetic impurity \cite{Chung96, Bulla08, Satori92, Sakai93}. Quantum fluctuations of the polarization of a magnetic impurity lead to the Kondo effect, which renormalizes the exchange interaction between the impurity and itinerant electrons at low energies. The renormalized interaction is expressed in terms of the Kondo temperature, $T_K\propto\exp{(-1/JN_0)}$ \cite{Balatsky06}. At small Kondo temperatures, $T_K\lesssim\Delta$, the effective constant $\beta\approx 1/\ln(\Delta/T_K)\lesssim 1$. Thus, whether in the ``classical" or ``quantum" consideration, the YSR level can be tuned arbitrarily close to the middle of the gap by an appropriate increase of the exchange constant.


The characteristic wavefunction of the YSR state is localized around the magnetic impurity and takes the form, $\phi_{sh}(\text{\bf r}) \sim \frac{1}{\text{\bf r}} e^{-\text{\bf r}/\xi |\sin(2\delta)|}$ \cite{Balatsky06}.  
For two impurities separated by distances $r \gg \xi$, the overlap between their associated YSR states is exponentially suppressed. 
However, for distances $r < \xi$, the YSR states of the two impurities hybridize, causing both an overall energy shift $\eta$ and a splitting $\eta'$, as depicted in Fig.~1b. 
Crucially, the overall energy shift $\eta$ depends on whether the impurity spins are aligned or anti-aligned; in particular, only in the anti-aligned case is it possible for a pair of YSR states to become virtually occupied by a Cooper pair from the superconducting condensate. This provides a natural intuition for our result: The  effective spin-spin interaction manifests as a consequence of the spin-dependence in  $\eta$.





With this intuition in mind, we now begin by considering the total energy associated with a pair of magnetic impurities (located at $r_L$ and $r_R$) in a superconductor. We treat the impurities as classical spins parallel to the $\hat{z}$ axis (which defines the direction in which the impurities are either aligned or anti-aligned). 
 The interaction Hamiltonian between the localized impurity and the itinerant electrons is then given by 
\begin{align}
H_{int} &= J \sum_{\sigma}   \int d \text{\bf r}  \sigma[ S_L  f(\text{\bf r} -\text{\bf r} _L) c_{\sigma}^{\dagger}(\text{\bf r} ) c_{\sigma} (\text{\bf r} ) \nonumber \\
 &+S_R  f(\text{\bf r} -\text{\bf r} _R) c_{\sigma}^{\dagger}(\text{\bf r} ) c_{\sigma} (\text{\bf r} )],     
\end{align}
where $S_{L(R)}$ is the spin of the left (right) impurity and $f(\text{\bf r} )$ characterizes the spatial form of the impurity potential. In momentum space, 
$H_{int} = J \sum_{\sigma} \int d{\bf \text{{\bf k}}} d{\bf \text{{\bf k}}}'   \sigma [ S_L e^{i (\text{\bf k} -\text{\bf k}')r_L} \tilde{f}_{\text{\bf k} ,\text{\bf k}'} + S_R e^{i (\text{\bf k}-\text{\bf k}')r_R} \tilde{f}_{\text{\bf k},\text{\bf k}'}]c_{\sigma,\text{\bf k}}^{\dagger} c_{\sigma,\text{\bf k}'} $, where $\tilde{f}$ is the Fourier transform of the potential. As is conventional \cite{Schrieffer}, we now define a Nambu spinor, $\Psi_\text{\bf k}=(c_{\ua,\text{\bf k}},c_{\da,-\text{\bf k}}^\dagger)$, wherein, 
$H_{0}=\int d\text{\bf k}  \Psi_{\text{\bf k}}^\dagger \left [ \epsilon_\text{\bf k}\tau^{z}+\Delta\tau^{x}  \right ] \Psi_{\text{\bf k}}$ ($\tau$ are Pauli matrices acting in particle-hole space). 
Similarly, the interaction becomes,
\begin{align}
H_{int} &= J  \int d\text{\bf k} d\text{\bf k}' \Psi_{\text{\bf k}}^\dagger [ S_L e^{i (\text{\bf k}-\text{\bf k}')r_L} \tilde{f}_{\text{\bf k},\text{\bf k}'}  \nonumber \\
&+ S_R e^{i (\text{\bf k}-\text{\bf k}')r_R} \tilde{f}_{\text{\bf k},\text{\bf k}'} ] \Psi_{\text{\bf k}'} + E_0
\end{align}
where $E_0 = -J  \int d\text{\bf k} \tilde{f}_{\text{\bf k},\text{\bf k}} [ S_L + S_R  ] $ arises from anti-commutation. 

Combining the bare BCS Hamiltonian and the interactions yields,
$H_{T} =H_0 + H_{int}$, which we  diagonalize utilizing a Bogoliubov transformation,
$d_n^\dagger= \int d\text{\bf k} (u_{n,\text{\bf k}}\psi^\dagger_{\ua,\text{\bf k}}+ v_{n,\text{\bf k}}\psi^\dagger_{\da,\text{\bf k}})$,  yielding,
\begin{equation}
H_{T}=\sum_n\varepsilon_n d_n^\dagger d_n-\frac{1}{2}\sum_n \varepsilon_n = \sum_n \varepsilon_n (d_n^\dagger d_n-\frac{1}{2}).
\end{equation}
The total energy of the ground state is thus given by
\begin{align}
E_{tot}=-\frac{1}{2}\sum_{n} |\varepsilon_n| = E_V -\frac{1}{2} \int d\epsilon |\epsilon| \delta\rho(\epsilon)
\end{align}
where $E_V$ characterizes the energy of the system in the absence of an impurity. Here,  $\delta\rho(\epsilon)$ represents the change in the total density of states as a result of the impurities and includes contributions from both continuum electronic states above the gap well as the discrete YSR states.  The effective exchange interaction, $I( {\bf r})$, between two impurities can be expressed in terms of changes to the DOS depending on whether the impurities are aligned or anti-aligned,
\begin{align}
I( {\bf r})=E_{tot}^{\uparrow, \downarrow} -E_{tot}^{\uparrow, \uparrow} =  -\frac{1}{2} \int d\epsilon |\epsilon| \left [ \delta \rho_{\uparrow, \downarrow}(\epsilon) - \delta \rho_{\uparrow, \uparrow}(\epsilon) \right ].
\end{align}

To calculate changes in the DOS, we compute $ \delta\rho(\epsilon) =- \frac{1}{\pi} \text{Im}\{\text{Tr}[G_{\text{\bf k},\text{\bf k}'}(z)-G^{(0)}_{\text{\bf k}}(z)] \} $,  where $z= \epsilon + i 0^{+}$, $G_{\text{\bf k}}^{(0)}(z) = [z - (\epsilon_\text{\bf k}\tau^{z}+\Delta\tau^{x} ) ]^{-1}$ is the bare BCS Green's function, and $G_{\text{\bf k},\text{\bf k}'}(z)$ is the perturbed Green's function. Since translational invariance is broken by the magnetic impurities, the perturbed Green's function depends on two momenta, \text{\bf k} and \text{\bf k}'.
Working within the $T$-matrix formalism \cite{Balatsky06}, 
\begin{align}
G_{\text{\bf k} ,\text{\bf k}'}(z) = G_{\text{\bf k}}^{(0)}(z) + G_{\text{\bf k}}^{(0)}(z) T_{\text{\bf k} ,\text{\bf k}'}G_{\text{\bf k}'}^{(0)}(z),
\end{align}
where  $T_{\text{\bf k} ,\text{\bf k}'}$ is the $T$-matrix.
Applying a Dyson  expansion to the $T$-matrix \cite{SuppInfo}, one finds that
\begin{align}
 \delta\rho(\epsilon) &=- \frac{1}{\pi} \text{Im} \{ \text{Tr}[G_{\text{\bf k}}^{(0)}(z) T_{\text{\bf k} ,\text{\bf k}'}G_{\text{\bf k}'}^{(0)}(z) ] \} \nonumber \\
 &=- \frac{1}{\pi} \text{Im} \{ \text{Tr}[JS\Pi (1-JSG)^{-1}] \}
\end{align}
where $\Pi$, $G$ and $S$ are $4\times 4$ matrices (in the tensor product space of particle-hole and left-right position) given by,
\begin{align}
&\Pi_{ll'} (z)=\int d\text{\bf k} G^{(0)}_\text{\bf k}(z)G^{(0)}_\text{\bf k}(z) e^{i  \text{\bf k} ( \text{\bf r}_l -  \text{\bf r}_{l'}) }       \\
&G_{ll'}(z)=\int d\text{\bf k} G^{(0)}_\text{\bf k}(z) e^{i  \text{\bf k} ( \text{\bf r}_l -  \text{\bf r}_{l'}) }    \\
&S_{ll'}=S_l\delta_{ll'} \otimes \tau^0.
\end{align}
Here, $\tau^0$ represents the identity matrix in particle-hole space and $l$, $l'$ run over $\{L, R\}$, indexing the left/right impurity.
%

We begin by considering the case of weakly bound YSR states ($J \ll 1$) and expand Eq.~(9) to second order in the exchange coupling, $\text{Tr}[JS\Pi (1-JSG)^{-1}]  \approx \text{Tr}[J^2S\Pi SG]$. 
Evaluating this perturbative expression results in the following superconducting RKKY exchange between the magnetic impurities, 
\begin{align}
I(\text{\bf r} )
 &= \frac{E_f \beta^2}{\pi (k_f r)^3}  \cos (2k_f r)e^{-\frac{2r  }{\xi}} F_1 \left [ \frac{2  r}{\xi} \right ] \nonumber \\
 &+  \frac{ \Delta \beta^2}{ (k_f r)^2}  \sin^2(k_fr)e^{-\frac{2 r }{\xi}} F_2 \left [ \frac{2 r }{\xi} \right ].
 \end{align} 
Here, $k_f$ is the Fermi momentum, $r = |\text{\bf r}_L -\text{\bf r}_R|$ is the distance between the spins and $
F_1 \left [ \alpha \right ] =  \alpha \int_0^{\infty}  dx e^{-\alpha (\sqrt{x^2+1}-1)}$, $F_2 \left [ \alpha \right ] =  \frac{2}{\pi} \int_0^{\infty}  dx \frac{e^{-\alpha( \sqrt{x^2 + 1}-1)} }     { (x^2 + 1)}$ are dimensionless integrals.  
The first term represents the bare RKKY interaction, while the second represents the antiferromagnetic correction resulting from superconductivity. Although this second term scales as $1/r^2$, it is weaker by a factor of $\Delta/E_f$ and only dominates over bare RKKY  at distances $r\gg E_f/(\Delta k_f)\sim\xi$, by which time the entire exchange integral $I(\text{\bf r})$ is exponentially suppressed. 
The above perturbative result is consistent with previous calculations which utilize the Kubo formula to compute the exchange interaction from the magnetization response  \cite{Galitski02, Aristov97, Abrikosov88, SuppInfo}. 
%
%

Returning to the interpretation of the exchange energy in terms of changes to the density of states [Eqs.~(7,9)], we recall that the effective exchange  contains two contributions, one from continuum electronic states and the other from  discrete YSR states. 
One might expect that, being only weakly bound, the YSR states should induce a contribution which decays more slowly than $e^{-\frac{2r }{\xi}}$.  However, we find that at $\mathcal{O}(J^2)$, the tail of the YSR contribution exactly cancels with a portion of the continuum contribution to yield the perturbative expression found in Eq.~(13).

\begin{figure}
\includegraphics[width=3.3in]{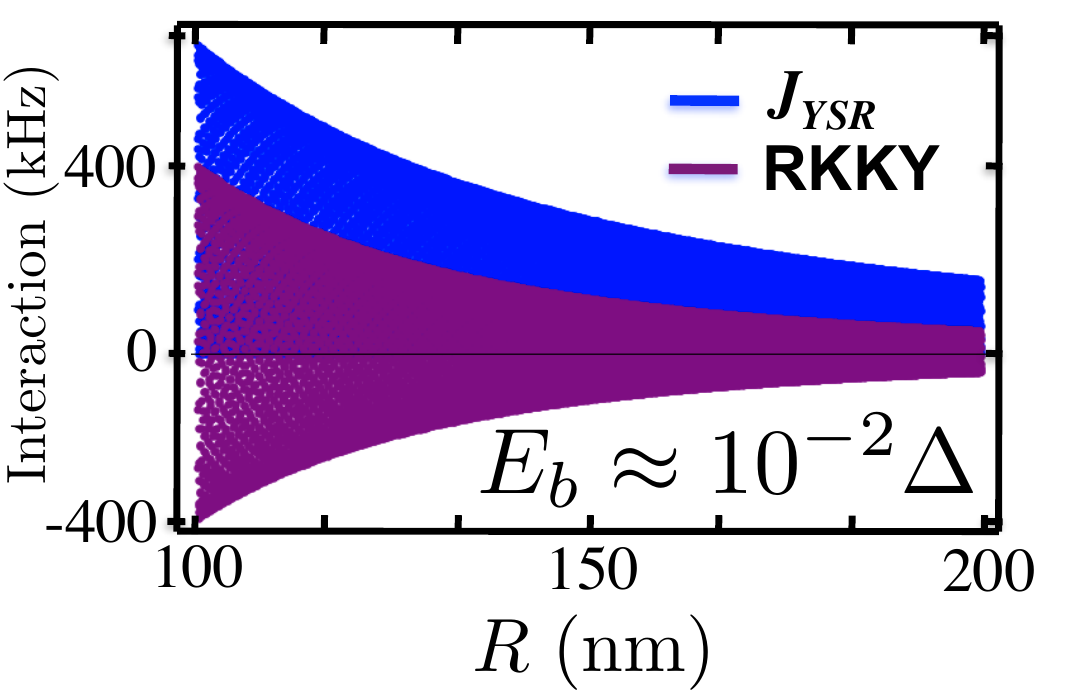}
\caption{(color online) For concreteness, all plots are calculated using actual parameters for superconducting Aluminum, with $E_f = 11.7$eV, $k_f =20.1$nm$^{-1}$, $N_0 = 35$eV/nm$^3$, and $\xi = 1.6\mu$m \cite{Donnelly81}.
 Comparison between bare RKKY and the  $J_{YSR}$  for $E_{b} \sim 10^{-2} \Delta$. Resonant enhancement  enables $J_{YSR}$ to dominate at distances $r \ll \xi$.  }
\label{fig:YSRscheme}
\end{figure}

Moving beyond the perturbative limit, as $J$ increases, the energy of the YSR bound state decreases (approaching the middle of the superconducting gap) and the relative strength of the continuum and YSR contributions change.
 In particular, one might expect the YSR contribution to dominate for deeply bound states for two reasons:  First, modifications to the bulk DOS will become weaker (since the bound state is further from the bottom of the band), and second, YSR hybridization with the superconducting condensate will become stronger as $E_b \rightarrow 0$.  This second point suggests that the energy shift $\eta$ has the potential to develop a singular contribution, arising from the $|\epsilon|$ in Eq.~(6) near $\epsilon \approx 0$; thus, any singular contribution to the exchange interaction can only arise from the low energy YSR states.
 
To see these effects explicitly, we now compute the bound state energies as a function of impurity separation. This corresponds to a direct calculation of the discrete YSR contribution to Eq.~(7).  The  YSR bound state energies can be computed from  poles of  $\text{Tr}[G_{\text{\bf k} ,\text{\bf k}'}(z)] $.  More explicitly, $E_{b}$ is determined by 
\begin{align}
F(E_{b}) \equiv \text{Det}[1-SG(E_{b})] =0.
\end{align}
In the limit, $k_f r \gg 1$, one can consider the hybridization of the isolated YSR bound states to obtain perturbative corrections to the YSR energies.
We derive an analytic approximation for solutions of Eq.~(14) in the case of both parallel and anti-parallel impurities \cite{SuppInfo}.  By subtracting the bare YSR energy [Eq.~(2)], this allows us to compute the spin-dependent total energy shift $\eta$. Our perturbative expansion is in the parameter $\eta / E_b$ and remains valid so long as the energy shift is small relative to the bare YSR energy (see Eq.~(16) and below for a discussion of validity). 

We first consider the case of anti-parallel impurities where symmetry allows us to directly expand around the bare YSR energy, $F(E_{b}) +\eta_{\uparrow \downarrow} F'(E_{b}) =0 $. 
 A straightforward but tedious calculation then yields the leading term in $1-\beta$ as
$
\eta_{\uparrow \downarrow}  = \Delta \frac{1}{1-\beta} \frac{\cos^{2}(k_fr)}{2(k_f r)^2}   e^{-\frac{2r}{\xi}}.
$
In the case of parallel spins, the situation is slightly more complicated since one must extract the  total shift by averaging the split energies (Fig.~1b). This requires expanding to third order, $F(E_{b}) +\eta_{\uparrow \uparrow} F'(E_{b}) + \frac{1}{2}\eta_{\uparrow \uparrow}^2 F''(E_{b}) +  \frac{1}{6}\eta_{\uparrow \uparrow}^3 F'''(E_{b}) =0 $ and results in a non-singular shift, 
$\eta_{\uparrow \uparrow} = -\frac{\Delta}{2} \frac{\cos(k_fr)}{(k_fr)^2}e^{-\frac{2r}{\xi}}$, as $\beta \rightarrow 1$ \cite{SuppInfo}. 

The YSR contribution to the exchange, $I ({\bf r})$, is given by $J_{YSR} =\eta_{\uparrow \downarrow} -  \eta_{\uparrow \uparrow}$ \cite{SuppInfo}. 
Crucially, as the bound state energy approaches the middle of the superconducting gap ($E_b \rightarrow 0$, $\beta \rightarrow 1$), $J_{YSR}$ is dominated by the singular contribution in $\eta_{\uparrow \downarrow}$ yielding,
\begin{align}
J_{YSR} = \Delta \frac{1}{1-\beta} \frac{\cos^{2}(k_fr)}{2(k_f r)^2}   e^{-\frac{2r}{\xi}},
 \end{align}
 which exhibits a resonant enhancement of the form $\frac{1}{1-\beta}$. 
This resonant enhancement has an  intuitive explanation.  It arises from the hybridization of a pair of YSR states with the superconducting condensate; more specifically, when the impurities are anti-aligned, this hybridization occurs as a result of the conversion of a Cooper pair from the condensate into a pair of electrons in the YSR states. 
 Heuristically, this coupling to the condensate takes the form $\Delta U(r) c_{L, \uparrow}^{\dagger} c_{R, \downarrow}^{\dagger}$, where $U(r) = \cos(k_fr)/(k_fr)$ characterizes the  overlap between the bound states. While the ground state energy correction stemming from this coupling is generally suppressed by an energy denominator $2E_{b}$, as $\beta$ approaches unity, $E_b$ approaches zero, leading to the observed resonant enhancement.

The physical limit of the enhancement of this purely antiferromagnetic contribution is set by the condition that the YSR energies have not crossed zero, which in effect, would signify a parity changing transition. This condition also represents the regime of validity for $J_{YSR}$ as derived from the expansion of Eq.~(14). 
In combination with the constraint that $J_{YSR}$ dominates over bare RKKY interactions, 
we obtain a double-sided inequality,
\begin{equation}
k_f r  > \frac{1}{1-\beta} >  \frac{\xi}{r}.
\end{equation}
By stark contrast to the perturbative limit, where the superconducting correction dominates only at distances $r \gg \xi$, here, we find that the anti-ferromagnetic $J_{YSR}$ exchange can prevail at $r \sim \sqrt{\lambda_f \xi} \ll \xi$ and reaches a maximum ($ \sim \Delta / \sqrt{k_f \xi}$) at such distances.

\emph{Discussion---} Inspection reveals that the YSR-induced interaction strength, $J_{YSR} =  \eta_{\uparrow \downarrow}- \eta_{\uparrow \uparrow} $ scales as $\sim\frac{1}{r^2}$, exhibiting a weaker decay than conventional metallic RKKY interactions. We note that this power-law is in agreement with the perturbative superconducting correction in Eq.~(13);  as expected, for small $\beta$, our full non-pertubative calculation matches the perturbative results  \cite{SuppInfo}.  In comparison to bare RKKY interactions, one important qualitative observation is that, while oscillatory in nature, $J_{YSR}$ does not vary between ferromagnetic and antiferromagnetic couplings.   The antiferromagnetic nature of the superconducting YSR correction results from the fact that coupling to the condensate  occur most effectively for  anti-aligned impurities.  

For small impurity separation and weakly bound YSR states, the magnitude of the RKKY interaction dominates over $J_{YSR}$. However, as illustrated in Fig.~2, for bound state energies close to the middle of the gap,  resonant enhancement enables $J_{YSR} > J_{RKKY}$ at distances well below the coherence length; the dominance of this anti-ferromagnetic exchange is further highlighted by the weaker power-law decay as a function of $r$.
This effect will be especially pronounced for superconductors with relatively large coherence lengths.

To observe/utilize  the resonant enhancement of $J_{YSR}$ requires a system where the coupling strength between the impurity spin and the superconductor can be 
tuned continuously. In principle, any low-density system with a 
tunable DOS can provide a natural mechanism for controlling the exchange constant via a gate voltage. 
An example of such a scenario is found in graphene \cite{Chen11}, where the exchange coupling of magnetic defects can be altered by simply changing the carrier density. In combination with demonstrations of proximity-induced superconductivity \cite{Heersche07}, this suggests that graphene in contact with a superconductor may represent a promising system with which to realize tunable-energy YSR states. Such a system naturally possesses a large coherence length since the Fermi velocity remains substantial even at low carrier densities.
Interestingly, it may also be possible to further enhance the effects of an applied gate voltage by separating
the graphene from the superconductor via a layer of semiconductor such as
MoS$_2$ \cite{Mak10,Radisavljevic11}.


In summary, working beyond the Born approximation, we have derived an enhanced anti-ferromagnetic exchange between magnetic impurities on the surface of a superconductor. This interaction is intimately related to the existence of a single mid-gap bound Yu-Shiba-Rusinov (YSR) state near a  magnetic impurity; indeed, it is the hybridization of these YSR states, which induces a long range antiferromagnetic interaction between spin impurities. Although our results are formulated within the treatment of classical spins, such a description is consistent for high-spin magnetic ions such as those currently used in experiments (e.g.~Gd,~Mn,~Cr) \cite{Yazdani97, Ji08}. Finally, our work motivates  numerical renormalization group (NRG) studies, where quantum fluctuations can be accounted for, shedding light on the interplay between Kondo singlet formation and YSR-induced spin-spin interactions \cite{Gerge}.

It is a pleasure to gratefully acknowledge the insights of and discussions with Gergely Zarand, Falko Pientka, Gang Chen, John Preskill,  Senthil Todadri, V. Galitski and Felix von Oppen. This work was supported, in part, by the NSF, DOE (FG02-97ER25308), HQOC Fellowship and DARPA at Harvard and the DOE Contract No.~DE-FG02-08ER46482 at Yale.

\end{document}


\title{Supplemental Material for

Enhanced anti-ferromagnetic exchange between magnetic impurities in a superconducting host}
\maketitle

\section{Perturbative Superconducting Correction to RKKY}

\noindent We provide the derivation for the perturbative (second order) superconducting correction to RKKY. In particular, we compute the exchange integral from the magnetization density,
\begin{align}
I(r) = \frac{1}{4\pi} \int_0^{\infty} \text{Tr}[ G_0(r;z) G_0(r;z)] dx
\end{align}
where $G_0(r;z) = \int d^3 k  \frac{z + \epsilon \tau_z + \Delta \tau_x}{z^2 - \epsilon^2 - \Delta^2} $ is the superconducting Green's function. Calculation reveals,
\begin{align}
G_0(r;z=ix) &=   \frac{2 \pi  \rho_0}{k_f r}   \frac{e^{ -\sqrt{\Delta^2+x^2}/v_f r }  }{  \sqrt{\Delta^2+x^2} }  \left (  \cos (k_f r)  \sqrt{\Delta^2+x^2}\tau_z    + \sin(k_f r)  [ ix  + \Delta \tau_x ] \right).
\end{align}
Computing $\text{Tr}[ G_0(r;z) G_0(r;z)]$ one finds,
\begin{align}
I(r) &= \int_0^{\infty} dx \frac{e^{-\frac{k_f r_0 \sqrt{x^2 + \Delta^2}}{E_f}}  \beta^2 [\Delta^2 +x^2 \cos (2k_f r_0)] }{\pi (k_f r_0)^2 (x^2 + \Delta^2)} \\
&= \frac{ \beta^2}{\pi (k_f r_0)^2} \left (  \cos (2k_f r_0) \int_0^{\infty} dx \left  [ e^{-\frac{k_f r_0 \sqrt{x^2 + \Delta^2}}{E_f}} \right   ] + 2\Delta^2 \sin^2(k_fr) \int_0^{\infty} dx \frac{e^{-\frac{k_f r_0 \sqrt{x^2 + \Delta^2}}{E_f}} }{ (x^2 + \Delta^2)} \right) 
\end{align} 
Changing the  variables from $x \rightarrow \Delta x$ yields, 
\begin{align}
I(r)&= \frac{ \beta^2}{\pi (k_f r_0)^2} \left (  \cos (2k_f r_0) \int_0^{\infty} \Delta dx \left  [ e^{-\frac{k_f r_0 \Delta \sqrt{x^2 + 1}}{E_f}} \right   ] + 2 \sin^2(k_fr) \int_0^{\infty} \Delta dx \frac{e^{-\frac{k_f r_0 \Delta \sqrt{x^2 + 1}}{E_f}} }{ (x^2 + 1)} \right) 
\end{align} 
Noting that
$ \int_1^{\infty}  dx \left  [ e^{-\frac{k_f r_0 \Delta x}{E_f}} \right   ] =  \frac{E_f}{k_f r_0 \Delta}  e^{-\frac{k_f r_0 \Delta }{E_f}}$, we can rewrite exactly,
\begin{align}
\int_0^{\infty}  dx \left  [ e^{-\frac{k_f r_0 \Delta \sqrt{x^2 + 1}}{E_f}} \right   ] = \frac{E_f}{k_f r_0 \Delta}  e^{-\frac{k_f r_0 \Delta }{E_f}} F_1 \left [ \frac{k_f r_0 \Delta}{E_f} \right ]
\end{align} 
where $F_1$ is the integral defined as,
\begin{align}
F_1 \left [ \alpha \right ] =  \alpha \int_0^{\infty}  dx e^{-\alpha (\sqrt{x^2+1}-1)}.
\end{align} 
The second integral can be recast in a similar fashion, 
\begin{align}
 \int_0^{\infty}  dx \frac{e^{-\frac{k_f r_0 \Delta \sqrt{x^2 + 1}}{E_f}} }{ (x^2 + 1)} = \frac{\pi}{2}e^{-\frac{k_f r_0 \Delta }{E_f}} F_2 \left [ \frac{k_f r_0 \Delta}{E_f} \right ]
\end{align} 
where $F_2$ is the integral defined as,
\begin{align}
F_2 \left [ \alpha \right ] =  \frac{2}{\pi} \int_0^{\infty}  dx \frac{e^{-\alpha( \sqrt{x^2 + 1}-1)} }     { (x^2 + 1)}.
\end{align} 
In combination, this yields the perturbative superconducting RKKY exchange (see Fig.~S1 for agreement with full non-perturbative calculation) as,
\begin{align}
I(r)&= \frac{E_f \beta^2}{\pi (k_f r_0)^3}  \cos (2k_f r_0)e^{- \frac{k_f r_0 \Delta}{E_f}} F_1 \left [ \frac{k_f r_0 \Delta}{E_f} \right ]
 +  \frac{\Delta \beta^2}{ (k_f r_0)^2}  \sin^2(k_fr)e^{- \frac{k_f r_0 \Delta}{E_f}} F_2 \left [  \frac{k_f r_0 \Delta}{E_f} \right ] \nonumber \\
 &= \frac{E_f \beta^2}{\pi (k_f r_0)^3}  \cos (2k_f r_0)e^{-\frac{2r_0  }{\xi}} F_1 \left [ \frac{2 r_0 }{\xi} \right ]
 +  \frac{ \Delta \beta^2}{ (k_f r_0)^2}  \sin^2(k_fr)e^{-\frac{2 r_0 }{\xi}} F_2 \left [ \frac{2 r_0 }{\xi} \right ] 
\end{align} 
While the integrals $F_1$ and $F_2$ cannot be performed analytically, by looking at the asymptotic behavior, one can provide a reasonable closed form approximation,
\begin{align}
F_1 \left [ \alpha \right ] =  \alpha \int_0^{\infty}  dx e^{-\alpha (\sqrt{x^2+1}-1)} \approx 1.25 (\alpha +0.65)^{1/2}
\end{align} 
\begin{align}
F_2 \left [ \alpha \right ] =  \frac{2}{\pi} \int_0^{\infty}  dx \frac{e^{-\alpha( \sqrt{x^2 + 1}-1)} }     { (x^2 + 1)} \approx \frac{0.8}{ (\alpha +0.65)^{1/2}}.
\end{align} 

\begin{figure}
\includegraphics[width=3.4in]{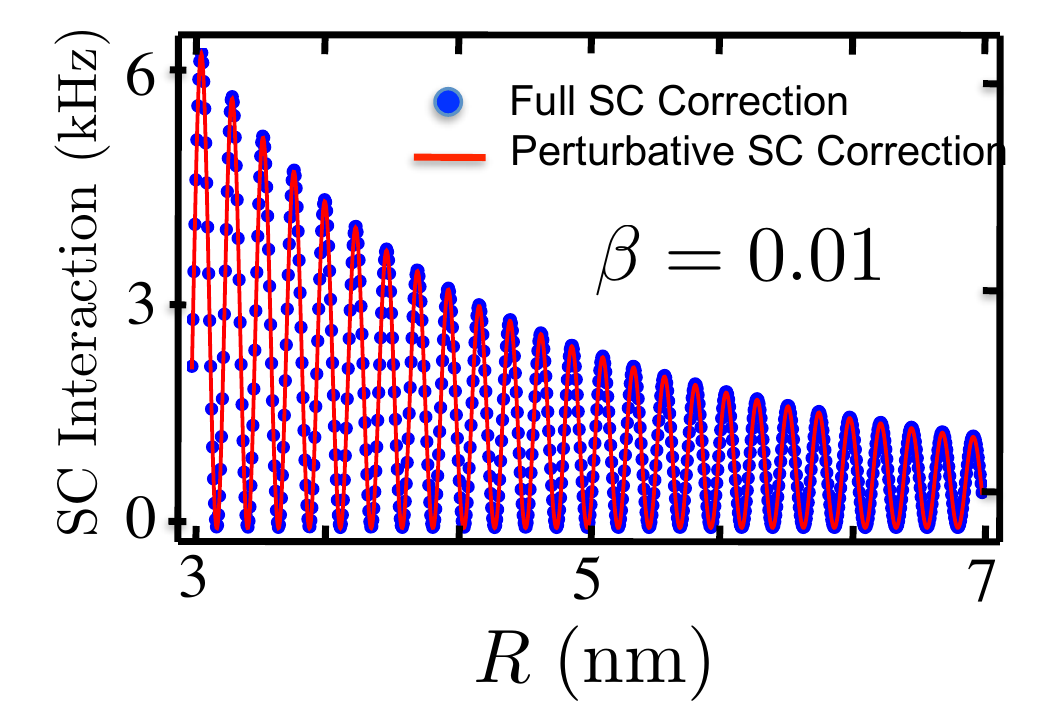}
\caption{Depicts a comparison between the non-perturbative SC correction (blue circles) and the perturbative correction (red line) for small $\beta= 0.01$. The non-perturbative SC corrections are obtained by numerically integrating the full Green's function and then subtracting out
the $\Delta= 0$ (metallic) portion.}
\label{fig:YSRscheme}
\end{figure}

\section{Dyson Expansion}

\noindent Recall from Eqn.~(9) of the main-text that the Dyson expansion reveals
\begin{align}
\text{Tr}[G_{\text{\bf k} ,\text{\bf k}'}(z)-G_{\text{\bf k}}^{(0)}(z) ] =\text{Tr}[S\Pi+S\Pi S G+S\Pi S G S G+\dots]=\text{Tr}[S\Pi(z) (1-S G^{(0)}(z))^{-1}].
\end{align}

\vspace{5mm}

\noindent We now establish that our above expression is consistent with the conventional linear response formalism in the small $S$ limit.
To check this, observe that at quadratic order in $S$ (dropping the linear order $S$ term, which does not contribute to interactions)
\begin{align}
\text{Tr}[S\Pi(z) (1-S G^{(0)}(z))^{-1}] \approx \text{Tr}[S\Pi(z)S G^{(0)}(z)].
\end{align}
Expanding both $\Pi(z)$ and $G(z)$ in eigenstates, we obtain 
\begin{align}
\text{Tr}[S\Pi(z)S G^{(0)}(z)]=\sum_{n,m} \frac{\langle\Psi_n|S|\Psi_m \rangle \langle \Psi_m|S|\Psi_n \rangle }{(z-\varepsilon_m)^2(z-\varepsilon_n)}.
\end{align}
The contribution of the above term to $E_{tot}$ involves terms of  the form 
\begin{align}
&\int_{0}^{\infty+i\Lambda}\frac{d\epsilon}{4\pi} 2 Re[\frac{i\epsilon}{(i\epsilon-\varepsilon_m)^2(i\epsilon-\varepsilon_n)}]\nonumber\\
&=\int_{-\infty-i\Lambda}^{\infty-i\Lambda}-\frac{\epsilon d\epsilon}{4\pi i}\frac{\epsilon}{(i\epsilon-\varepsilon_m)^2(i\epsilon-\varepsilon_n)}\nonumber\\
&=\int_{-\infty-i\Lambda}^{\infty-i\Lambda}-\frac{d\epsilon}{4\pi }\frac{\epsilon}{(\epsilon+i\varepsilon_m)^2(\epsilon+i\varepsilon_n)}.
\end{align}
The above integral can be computed via contour integration by closing the contour around either of $\pm i\infty$. If both $-i\varepsilon_{m,n}$ lie on the 
same side of $i\Lambda$ then the integral vanishes. Otherwise we can close the integral on the side of $i\Lambda$ containing $-i\varepsilon_n$ so that 
\begin{align}
&\int_{-\infty-i\Lambda}^{\infty-i\Lambda}-\frac{d\epsilon}{4\pi }\frac{\epsilon}{(\epsilon+i\varepsilon_m)^2(\epsilon+i\varepsilon_n)}\nonumber\\
&=-\frac{1}{2}\frac{\varepsilon_n\textrm{sgn}(\varepsilon_n-\Lambda)}{(\varepsilon_n-\varepsilon_m)^2}.
\end{align}
Symmetrizing with respect to $m$,  we obtain 
\begin{align}
&\int_{-\infty-i\Lambda}^{\infty-i\Lambda}-\frac{d\epsilon}{4\pi }\frac{\epsilon}{(\epsilon+i\varepsilon_m)^2(\epsilon+i\varepsilon_n)}\nonumber\\
&=-\frac{1}{4}\frac{\varepsilon_n\textrm{sgn}(\varepsilon_n-\Lambda)+\varepsilon_m\textrm{sgn}(\varepsilon_m-\Lambda)}{(\varepsilon_n-\varepsilon_m)^2}\\
&=-\frac{1}{4}\frac{\textrm{sgn}(\varepsilon_n-\Lambda)}{(\varepsilon_n-\varepsilon_m)}\\
&=\frac{1}{4}\int_{-\infty-i\Lambda}^{\infty-i\Lambda}-\frac{d\epsilon}{2\pi }\frac{1}{(\epsilon+i\varepsilon_m)(\epsilon+i\varepsilon_n)}
.
\end{align}
Substituting back, we obtain that the correction to the total energy is also given by 
\begin{align}
&4\delta E_{tot}=2\int_{0}^{\infty}\frac{d\epsilon}{4\pi} 2 \text{Re}[i  \text{Tr}[S G^{(0)}(i\epsilon)S G^{(0)}(i\epsilon)]] \nonumber \\
&-\int_{0}^{\infty+i\Lambda}\frac{d\epsilon}{4\pi} 2  \text{Re}[i  \text{Tr}[S G^{(0)}(i\epsilon)S G^{(0)}(i\epsilon)]]-\int_{0}^{\infty-i\Lambda}\frac{d\epsilon}{4\pi} 2  \text{Re}[i  \text{Tr}[S G^{(0)}(i\epsilon)S G^{(0)}(i\epsilon)]],
\end{align}
in agreement with the conventional linear response formalism.

\section{Expressions for $G_{LL}^{\sigma \sigma'} (z)  $,  $G_{LR}^{\sigma \sigma'} (z)  $, $\Pi_{LL}^{\sigma \sigma'} (z)  $ and $\Pi_{LR}^{\sigma \sigma'} (z)  $}

\noindent Here, we provide the derivations for $G_{LL}^{\sigma \sigma'} (z)  $, $G_{LR}^{\sigma \sigma'} (z)  $,$\Pi_{LL}^{\sigma \sigma'} (z)  $ and $\Pi_{LR}^{\sigma \sigma'} (z)  $ as used in the numerics. 
First, let us note that $L$ and $R$ will denote the left and right impurities. From the symmetry of the expressions, one can immediately see that $G_{LL}^{\sigma \sigma'} (z)  = G_{RR}^{\sigma \sigma'} (z)  $ and that $G_{LR}^{\sigma \sigma'} (z)  = G_{RL}^{\sigma \sigma'} (z)  $. Expressing the BCS Green's function yields,
\begin{align}
G_{LL}^{\sigma \sigma'} (z)  = \int d^3k \frac{1}{z^2 - \Delta^2 - \epsilon_k^2}  \left (
\begin{matrix} 
  z-\epsilon_k & \Delta   \\ 
  \Delta & z+\epsilon_k
\end{matrix} \right) =  \frac{-\rho_0} {\sqrt{\Delta^2-z^2}} \left (
\begin{matrix} 
  z& \Delta   \\ 
  \Delta & z
\end{matrix} \right)
\end{align}For  $G_{LR}^{\sigma \sigma'} (z)  $, the only difference is that the form factors do note cancel out. Instead, one finds 
\begin{align}
G_{LR}^{\sigma \sigma'} (z)  = \int d^3k \frac{1}{z^2 - \Delta^2 - \epsilon_k^2}  \left (
\begin{matrix} 
  z-\epsilon_k & \Delta   \\ 
  \Delta & z+\epsilon_k
\end{matrix} \right) e^{i k \cdot r_0},
\end{align}
where $\vec{r_0} = \vec{r_L} - \vec{r_R}$. Contour integration yields
\begin{align}
\hspace{-20mm}
G_{LR}^{\sigma \sigma'} (z)  = \frac{-\rho_0}{k_f r_0} \frac{  e^{- \frac{\sqrt{\Delta^2 - z^2}}{v_f}r_0}}{\sqrt{\Delta^2-z^2}}  \left (
\begin{matrix} 
  z \sin (k_f r_0) + \sqrt{\Delta^2- z^2} \cos (k_f r_0) & \Delta\sin (k_f r_0)   \\ 
  \Delta\sin (k_f r_0) & z \sin (k_f r_0) - \sqrt{\Delta^2- z^2} \cos (k_f r_0)
\end{matrix} \right).
\end{align}
A similar calculation for $\Pi_{LL}^{\sigma \sigma'} (z) $ yields,
\begin{align}
\Pi_{LL}^{\sigma \sigma'} (z)  =  \frac{\rho_0 \Delta} {(\Delta^2-z^2)^{3/2}} \left (
\begin{matrix} 
  \Delta& z   \\ 
  z & \Delta
\end{matrix} \right).
\end{align}
Finally, we turn to $\Pi_{LR}^{\sigma \sigma'} (z) $. Taking care to close the contours in the correct plane, we obtain,
\begin{align}
\hspace{-15mm}
\frac{\rho_0} {2E_f } \frac{e^{- \frac{\sqrt{\Delta^2 - z^2}}{v_f}r_0}}{(\Delta^2-z^2)^{3/2}} 
\left (
\begin{matrix} 
  z (\Delta^2-z^2) \cos (k_f r_0) + (\frac{2E_f}{k_f r_0} \Delta^2 +  z^2 \sqrt{ \Delta^2-z^2}) \sin (k_f r_0)& z\Delta \left (\frac{2E_f}{k_f r_0} + \sqrt{\Delta^2-z^2}\right) \sin (k_f r_0)   \\ 
  z\Delta \left (\frac{2E_f}{k_f r_0} + \sqrt{\Delta^2-z^2}\right) \sin (k_f r_0) & z (z^2-\Delta^2) \cos (k_f r_0)  + (\frac{2E_f}{k_f r_0} \Delta^2 +  z^2 \sqrt{ \Delta^2-z^2}) \sin (k_f r_0) 
\end{matrix} \right). \nonumber
\end{align}

\section{Singular YSR Contribution}

\noindent Here, we explain the dominance of the YSR contribution relative to the bulk contribution (e.g. changes to the bulk DOS). In the limit $k_fr \gg 1$, we derive an analytic expression for the YSR shifts in both the parallel and anti-parallel impurity configuration. 
\begin{align}
\hspace{-3mm}
\eta_{\uparrow \downarrow}  &= E_{sh} \frac{\tan^{2}(2\delta)}{(2k_f r)^2} \left [ \frac{1+\beta^2 + (3\beta^2-1)\cos(2k_fr)}{1+\beta^2} \right] e^{-\frac{2r}{\xi}\sin(2\delta)}  \\ 
&= \Delta \frac{1-\beta^2}{1+\beta^2} \frac{1}{(2k_f r)^2} \left ( \frac{2\beta}{1-\beta^2}  \right ) ^2\left [ \frac{1+\beta^2 + (3\beta^2-1)\cos(2k_fr)}{1+\beta^2} \right] e^{-\frac{2r}{\xi}\sin(2\delta)} \\
&= \Delta \frac{1}{1-\beta^2} \frac{1}{(2k_f r)^2}  \frac{4\beta^2}{1+\beta^2}  \left [ \frac{1+\beta^2 + (3\beta^2-1)\cos(2k_fr)}{1+\beta^2} \right] e^{-\frac{2r}{\xi}\sin(2\delta)} 
\end{align}
\begin{align}
\eta_{\uparrow \uparrow} &=  E_{sh} \frac{\sin^{2}(2\delta)}{(2k_f r)^2} \left [ \frac{-1+\beta^2 + (1-5\beta^2)\cos(2k_fr)}{1-\beta^2} \right.  \left. + \frac{2r}{\xi} \sin(2\delta) \sin^2 (k_fr) \right ] e^{-\frac{2r}{\xi}\sin(2\delta)} \\
&= \Delta \frac{1-\beta^2}{1+\beta^2} \frac{1}{(2k_f r)^2} \left( \frac{2\beta}{1+\beta^2} \right)^2 \left [ \frac{-1+\beta^2 + (1-5\beta^2)\cos(2k_fr)}{1-\beta^2} \right.  \left. + \frac{2r}{\xi} \frac{2\beta}{1+\beta^2}\sin^2 (k_fr) \right ] e^{-\frac{2r}{\xi}\sin(2\delta)} \\
&= \Delta \frac{1}{1+\beta^2} \frac{1}{(2k_f r)^2} \left( \frac{2\beta}{1+\beta^2} \right)^2 \left [ -1+\beta^2 + (1-5\beta^2)\cos(2k_fr) \right.  \left. + \frac{2r}{\xi} \frac{2\beta (1-\beta^2)}{1+\beta^2}\sin^2 (k_fr) \right ] e^{-\frac{2r}{\xi}\sin(2\delta)} 
\end{align}
As one can see the singular contribution to the YSR interaction comes from the divergence of $\eta_{\uparrow, \downarrow}$  as $\beta \rightarrow 1$ ($E_{sh} \rightarrow 0$). The form for this singular portion of the YSR interaction is given by,
\begin{align}
\hspace{-3mm}
J_{YSR}^{singular}= \Delta \frac{1}{1-\beta} \frac{\cos^{2}(k_fr)}{2(k_f r)^2}   e^{-\frac{2r}{\xi}} .
\end{align}